\documentclass[slac_one]{revtex4}
\usepackage{graphicx}
\usepackage{fancyhdr}
\begin{document}

\title{Heavy-light meson's physics in Lattice QCD} %% Paper title goes here


\author{N. Tantalo}
\affiliation{INFN sezione di Roma "Tor Vergata", Via della Ricerca Scientifica 1, 00133 Rome, Italy}
\affiliation{Centro Ricerche e Studi E. Fermi, Compendio Viminale, 00184 Rome, Italy}

\begin{abstract}
The possibility of revealing new physics by studying the flavor sector of 
the Standard Model strongly depends upon the accuracy that will be achieved 
in (near) future lattice QCD calculations and, in particular, on
heavy-light meson's observables.
In turn, handling with heavy-light mesons on the lattice is a challenging problem, because 
of the presence of two largely separated energy scales, and at present it is 
impossible to extract matrix elements involving B mesons in external states 
without recurring to some approximation.
In this note I give a fast overview of some of the methods that have been devised to handle such 
kind of problems, emphasizing those based on finite volume techniques, and briefly discuss some recent
results obtained by their application.
\end{abstract}

\maketitle

\thispagestyle{fancy}

%%%%%%%%%%%%%%%%%%%%%%%%%%%%%%%%%%%%%%%%%%%%%%%%%%%%%%%%%%%%%%%%%%%%%%%%%%%%%%%%%%%%%%%%%%%%%%%%%%%%%%%%%%%
%\section{INTRODUCTION}

Accurate and reliable theoretical calculations of many Standard Model processes are still lacking, 
particularly of the QCD contributions to decay rates and scattering amplitudes. 
Lattice QCD may eventually provide the required non-perturbative accuracy though
numerical calculations are inevitably affected by systematics errors. Calculations are performed at
finite lattice spacings, finite volumes and at unphysical values of the light quark masses but
reliable estimates of these systematics can be obtained by performing simulations of several values of the cutoff, different physical volumes and of several values of the running quark masses.
On a different ground have to be considered those systematics that cannot be reliably quantified like, for
example, quenching, rooting etc.
Quenching is not an issue anymore thank to several improvements 
that have been achieved in the field of simulation algorithms that opened
the way toward simulations of full QCD (including the effects of sea quarks) with
light quark masses falling within the range of applicability of chiral perturbation theory
(see refs.~\cite{Giusti:2007hk,jansenlat} for recent reviews).
Concerning rooted staggered fermions, a proof in favor or against the validity of the
"fourth root trick" is still lacking while, at the same time, several hadronic
observables have been calculated by using this formalism, ranging from $f_\pi$
to $M_\Upsilon$, and many of them are in fairly good agreement with available experimental
determinations. 
Several authors have discussed this issue from different points of view and I will not 
enter in further details here (see for example refs.~\cite{Sharpe:2006re,Creutz:2007rk,Kronfeld:2007ek}). 

When handling with heavy-light mesons on the lattice one is forced to introduce additional sources of
systematics because of the simultaneous presence of two largely separated energy scales, the
heavy quark mass and the confinement scale. 
A direct simulation of relativistic $b$-quarks would be possible by generating full QCD gauge ensembles
on lattices with a number of points of the order of a few hundred per spatial direction , i.e. by 
taking under control at the same time cutoff effects $O(am_h)$ (or $O(am_h)^2$ in improved theories) and 
finite volume effects $O(e^{-\Lambda_{QCD}L})$. 
Presently available super computers allow the simulation of lattices with about $50$ number of points per spatial direction (taking into account the necessity of continuum limit extrapolations)
and several strategies have been devised to cope with such kind of two-scale problems. 
Different strategies can be classified in "large volume" approaches (LVA) and "small volume" approaches 
(SVA) depending on which one of the two scales is properly accommodated on the lattice. 

By studying heavy-light mesons on lattices corresponding to physical volumes where
pions as light as $M_\pi\sim 300$~MeV can be simulated by meeting the bound $M_\pi L \gtrsim 3$, finite 
volume effects can be safely neglected in a first analysis since they are certainly smaller than the ones 
encountered in the study of light pseudoscalar mesons. 
On the contrary, having a limited number of lattice points per spatial direction, cutoff effects
coming from relativistic propagating $b$-quarks would be too large ($am_b\simeq 1$) and a proper handling of the heavy degrees of freedom requires the introduction of some approximation, typically effective field theories. 

A first possibility consists in simulating relativistic heavy quarks with
masses smaller or at best equal to the physical value of the charm quark mass and in extrapolating
the observables of interest to the beauty mass by relying on analytical formulae derived by means of
heavy quark effective theory (HQET). The extrapolations can be (and for several observables have been) 
turned into interpolations by introducing static quarks on the lattice through the Eichten-Hill
action~\cite{Eichten:1989zv}:
\begin{eqnarray}
\mathcal{L}_{HQET}=\bar{\psi}_h
(D_0+\delta m)\psi_h-\underbrace{\bar{\psi}_h\left(\omega_{spin}\ \vec{\sigma}\cdot\vec{B}
+\frac{\omega_{kin}}{2} \vec{D}^2\right)\psi_h+O(m_h^{-2})}_{insertions}, \qquad \psi_h=\frac{1+\gamma_0}{2}\ \psi
\end{eqnarray}
This action is renormalizable provided that the sub-leading terms ($\omega_X\propto m_h^{-1}$) are treated 
as insertions. Observables computed within the HQET framework need to be renormalized and matched to the
corresponding QCD quantities at a matching scale $\mu\simeq m_b$. If lattice simulations are performed
on large volumes, the matching with QCD can be only performed perturbatively thus compromising
the non-perturbative accuracy that a lattice calculation should provide. 
Furthermore, the renormalization procedure is 
particularly cumbersome on the lattice because of the presence of power divergences that arise
as a consequence of the mixing of operators of lower dimensions with the observable of interest. 
A solution to these problems has been found by Heitger and Sommer~\cite{Heitger:2003nj} and it is 
based on a finite volume technique. In the following I will just illustrate the basic idea behind this 
method remanding  to ref.~\cite{Sommer:2006sj} for an exhaustive review and to 
ref.~\cite{DellaMorte:2007ij} for a practical application of the method to the calculation of the leptonic decay constants of the heavy-strange mesons. 
The small volume approach to HQET consists in simulating lattices with spacings that allow the propagation 
of relativistic $b$-quarks with small cutoff effects ($am_b\lesssim 1/3$) at the price of 
introducing large finite volume effects. On such small volumes one computes a set of suitably chosen 
correlation functions both in QCD and in HQET and extracts the matching coefficients at the scale 
$\mu\simeq m_b$,
\begin{equation} 
\mathcal{O}^{QCD}(m_b;L_0) = C(m_b) \mathcal{O}^{HQET}(L_0)
\end{equation} 
where $\mathcal{O}$ is the observable, $L_0$ the physical extension of the volume, that in practical
application is of the order of $0.5$~fm, and $C(m_b)$ is the matching coefficient. Physical results
are subsequently obtained by performing simulations on progressively large volumes and by calculating, 
within the effective theory, the so-called "step scaling functions" (SSF), formally defined by
\begin{equation} 
\mathcal{O}^{HQET}(2L) = \sigma_{\mathcal{O}}(\mathcal{O}^{HQET}(L))
\end{equation} 
At each step of the calculation the continuum limit can and must be taken and at the end of the game
one ends up with the observable calculated within HQET, renormalized and matched with 
non-perturbative accuracy.

A second large volume approach consists in simulating heavy quarks on lattices with spacings such that
$am_b\sim 1$ by using the so-called Fermilab (FNAL) 
lagrangian~\cite{ElKhadra:1996mp,Aoki:2001ra,Oktay:2008ex},
\begin{equation} 
\mathcal{L}_{FNAL}=\bar{\psi}\left[m_0+\gamma_0 D_0+\zeta\vec{\gamma}\cdot \vec{D}
	  -r_t\frac{aD_0^2}{2}-r_s\frac{a\vec{D}^2}{2}
	  +c_B\frac{i\sigma_{ij}F_{ij}}{4}+c_E\frac{i\sigma_{0i}F_{0i}}{2}
	  \right]\psi
\end{equation} 
i.e. the Symanzik effective action for quarks at small spatial momenta ($\vert a\vec{p}\vert\ll 1$) 
with mass dependent coefficients. The improvement coefficients are calculated by matching FNAL observables
to QCD ones at $m_h\simeq m_b$. As in the previous case, by making simulations on large volumes
the matching can be only performed perturbatively. Recently it has been observed~\cite{Christ:2006us}
that the number of independent coefficients can be reduced to three and that they can be determined
non-perturbatively by making, again, simulations on a small physical volume but with a fine lattice
spacing~\cite{Lin:2006ur}.

By this fast (and incomplete) review of large volume approaches to heavy-light meson's physics
on the lattice it comes out that the preservation of (full) non-perturbative accuracy requires
simulations to be performed at fine lattice spacings ($am_b\ll 1$) and, for the time being, 
on small physical volumes. 

In ref.~\cite{Guagnelli:2002jd} we have proposed a different finite volume method to handle with
two-scale problems on the lattice, subsequently applied in refs.~\cite{deDivitiis:2003iy,deDivitiis:2003wy} 
to the quenched calculation of the $b$-quark mass and of the heavy-strange pseudoscalar leptonic
decay constants, the so-called "step scaling method" (SSM). 
Within the SSM approach the small volume calculations are needed in order to
resolve the dynamics of the heavy quarks without recurring to any approximation but
introducing, at intermediate stages, finite volume effects (FVE).
\begin{figure*}[t]
\centering
\includegraphics[width=0.45\textwidth]{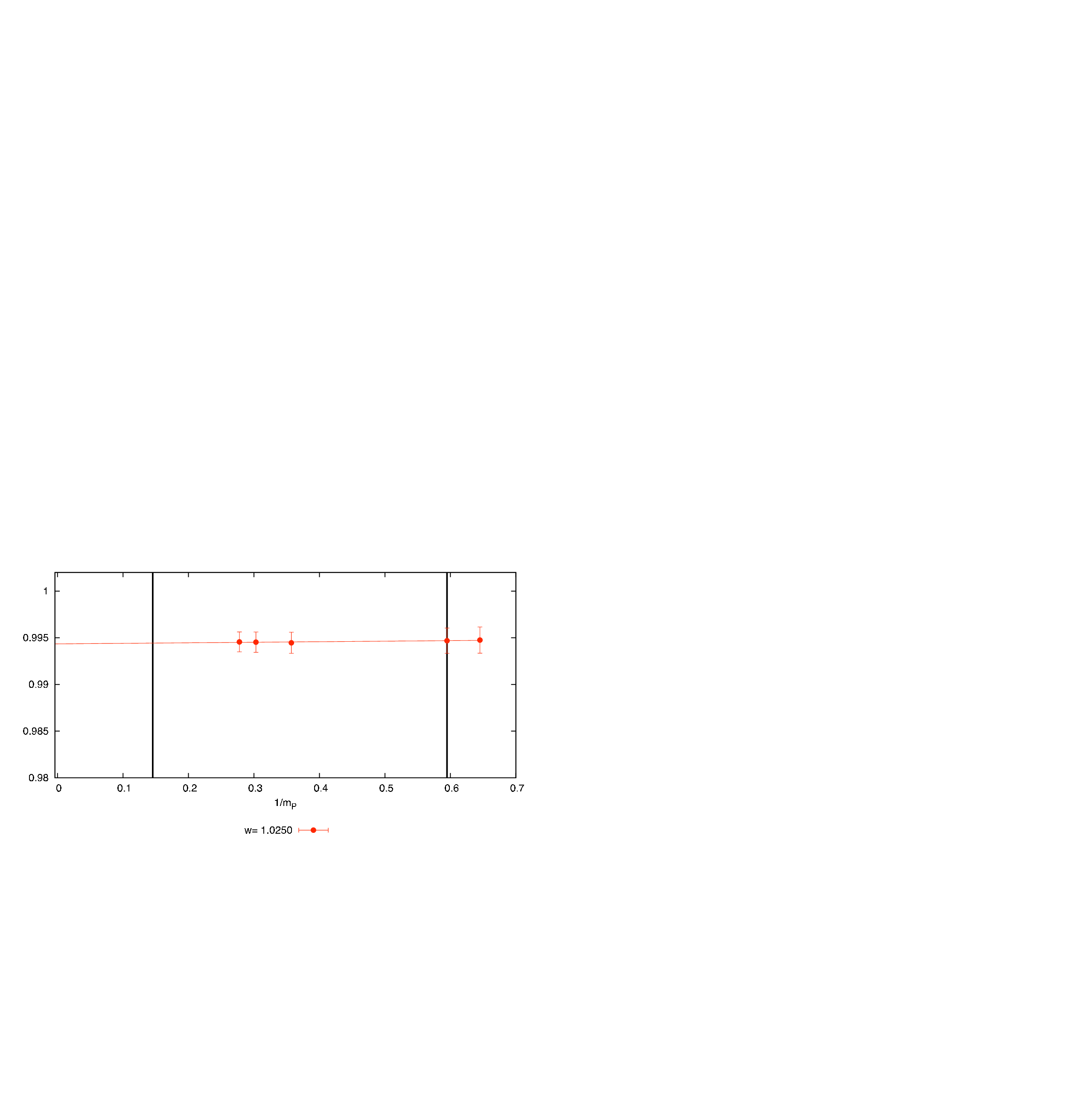}
\includegraphics[width=0.45\textwidth]{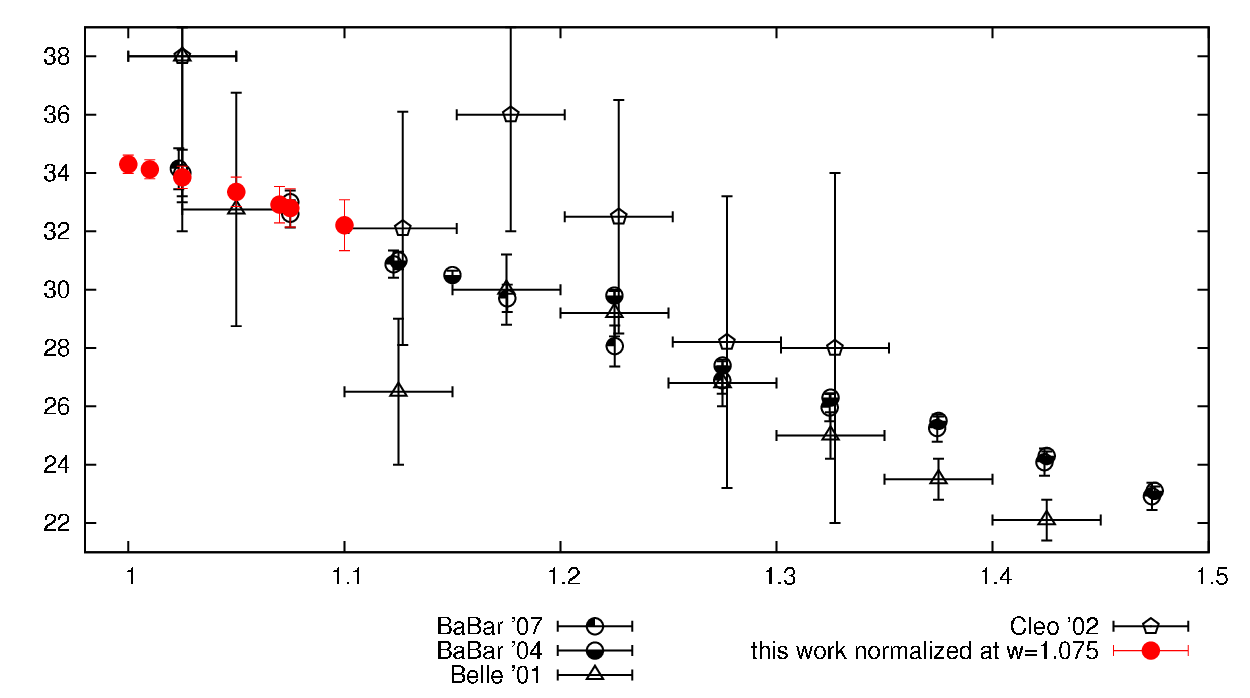}
\caption{The plot on the left shows the step scaling function 
$\sigma^{P\rightarrow D^{\star}}(w;L_0,L_1)$ at a fixed value of $w$ as a
function of the inverse mass of the heavy quark of the pseudoscalar meson;
the black vertical lines represent the physical values of the bottom and charm masses.
The plot on the right shows physical results
for $\vert V_{cb}\vert F^{B\rightarrow D^\star}(w)$ compared with some of the available 
experimental determinations.} \label{ssf_final}
\end{figure*}
\begin{eqnarray}
\mathcal{O}(physical)\; =\; \mathcal{O}(finite\ volume)\;\times\; FVE 
\end{eqnarray}
The finite volume effects are subsequently accounted for
by performing simulations on progressively larger volumes and by relying on the
observation that FVE are almost entirely due to the low energy scale. 
The success of this strategy depends on the details of the problem and hence on the possibility of
computing the finite volume observable, finite volume effects and their
product with smaller errors and systematics with respect to the
ones that would be obtained by a direct calculation. The strength of the
method is a great freedom in the definition of the observable on finite volumes
provided that its physical value is recovered at the end of the procedure. In order to explain
how the method works in practice I will specialize the formulae to the case of the form
factors 
$\mathcal{O}^{B\rightarrow D^{(\star)}}(w)=\{G^{B\rightarrow D}(\omega),F^{B\rightarrow D^\star}(w)\}$ entering the semileptonic decays $B\rightarrow D^{(\star)}\ell\nu$,
\begin{eqnarray}
\frac{d\Gamma(B\rightarrow D\ell \nu)}{d\omega} &=& (\mbox{kin. fact.}) \
\vert V_{cb} \vert^2
(\omega-1)^{\frac{3}{2}} \left[G^{B\rightarrow D}(\omega)\right]^2
\\ \nonumber \\ 
\frac{d\Gamma(B\rightarrow D^\star\ell\nu)}{dw}&=&(\mbox{kin. fact.}) \
\vert V_{cb}\vert^2 
(w-1)^{\frac{1}{2}}\left[F^{B\rightarrow D^\star}(w)\right]^2
\end{eqnarray}
where $w=p_i\cdot p_f/M_iM_f$. The proper definition of the form factors on the lattice
in terms of three-point correlation functions can be found in 
refs.~\cite{de Divitiis:2007uk,de Divitiis:2007ui,deDivitiis:2008df} where the calculations have been
carried out.
First we compute the observable $\mathcal{O}^{B\rightarrow D^{(\star)}}(w;L_0)$ on a small volume
$L_0\simeq 0.4$~fm, which is chosen to accommodate the dynamics of the $b$-quark.
A first portion of finite volume effects is removed by evolving the volume from $L_0$ to 
$L_1=2L_0$, by the ratio
\begin{eqnarray} 
\sigma^{P\rightarrow D^{(\star)}}(w;L_0,L_1)=\frac{\mathcal{O}^{P\rightarrow D^{(\star)}}(w;L_1)}
{\mathcal{O}^{P\rightarrow D^{(\star)}}(w;L_0)}
\label{eq:sigma1}
\end{eqnarray}
The crucial point is that the step scaling functions are calculated by simulating
heavy quark masses $m_P$ smaller than the $b$-quark mass. The physical
value $\sigma^{B\rightarrow D^{(\star)}}(w;L_0,L_1)$ is obtained by a smooth extrapolation in $1/m_P$ that
relies on the HQET expectations (see ref.~\cite{deDivitiis:2003iy}) 
and upon the general idea that finite volume effects, measured
by the $\sigma$'s, are almost insensitive to the high energy scale. 
The final result is obtained by further evolving the volume from $L_1$ 
to $L_2=4L_0$ by calculating a second step scaling function and by the following identity
\begin{eqnarray} 
\mathcal{O}^{B\rightarrow D^{(\star)}}(w;L_2)=
\mathcal{O}^{B\rightarrow D^{(\star)}}(w;L_0)
\;\sigma^{B\rightarrow D^{(\star)}}(w;L_0,L_1)
\; \sigma^{B\rightarrow D^{(\star)}}(w;L_1,L_2)
\label{eq:infresultformula}
\end{eqnarray}
How good is the hypothesis of low sensitivity of the FVE with respect to the high energy scale can
be appreciated by looking at the plot in the left panel of Figure~\ref{ssf_final} in which the
step scaling function $\sigma^{P\rightarrow D^{\star}}(w;L_0,L_1)$ is plotted at a fixed value of $w$ as a
function of the inverse mass of the heavy quark of the pseudoscalar meson. Similar plots are shown
for the other step scaling functions together with continuum extrapolations in
refs.~\cite{de Divitiis:2007uk,de Divitiis:2007ui,deDivitiis:2008df}. The physical results
for $\vert V_{cb}\vert F^{B\rightarrow D^\star}(w)$ are compared with some of the available 
experimental determinations in the plot in the right panel of Figure~\ref{ssf_final}.
These results have been obtained within the quenched approximation with the aim of checking the validity
of the method and to explore the feasibility of a future full QCD calculation.
Nevertheless they may have some phenomenological relevance since they allow the extraction of $V_{cb}$ 
without extrapolating experimental data. In fact no other results are presently
available for the form factors beyond the point at zero recoil ($w\ge1$).

The step scaling method can be combined with the finite volume approach to HQET in order to turn
the already mild extrapolations of the step scaling functions into interpolations and further
improve the accuracy of the results~\cite{Guazzini:2007ja}. So far, these two methods have been used
only in quenched applications. Unquenched results for the leptonic decay constants of the heavy-light
mesons, the $b$-quark mass and the semileptonic form factors will be produced in the near future by
using the large physical volumes gauge ensembles generated by the Coordinated Lattice Simulations
(CLS) effort and by producing small volume full QCD gauge ensembles satisfying Schr\"odinger Functional 
boundary conditions~\cite{DellaMorte:2007qw}. Some preliminary results have already been obtained
and can be found in refs.~\cite{vonHippel:2008pc,DellaMorte:2008du} together with a presentation of 
the CLS initiative.

%%%%%%%%%%%%%%%%%%%%%%%%%%%%%%%%%%%%%%%%%%%%%%%%%%%%%%%%%%%%%%%%%%%%%%%%%%%%%%%%%%%%%%%%%%%%%%%%%%%%%%%%%%%
\begin{acknowledgments}
I thank the organizers of the ICHEP conference for the kind hospitality in Philadelphia. 
A warm thank goes to G.M. de Divitiis
and R. Petronzio for a friendly and fruitful collaboration during the last years on the subjects
covered into this talk. A special thank goes also to R. Sommer.
\end{acknowledgments}

%%%%%%%%%%%%%%%%%%%%%%%%%%%%%%%%%%%%%%%%%%%%%%%%%%%%%%%%%%%%%%%%%%%%%%%%%%%%%%%%%%%%%%%%%%%%%%%%%%%%%%%%%%%

\end{document}